\documentclass[journal=nalefd,manuscript=letter ]{achemso}

\usepackage{graphicx,epsfig,amsmath, amssymb, multicol, setspace}
\usepackage[font=footnotesize]{caption}

\newcommand{\fig}[1]{~\ref{fig:#1}}
\newcommand{\eqn}[1]{~\ref{eq:#1}}
\newcommand{\Vds}{V$_{ds}$}
\newcommand{\Vg}{V$_{g}$}

\title{Charge dynamics of individual conductance channels within a percolation network of a nano-patterned nanocrystal quantum dot solid }
\author{Xiangxi Yin} 
\affiliation{Department of Materials Science and Engineering, University of California Riverside, Riverside, California 92521}
\author{Bence Papp} 
\affiliation{Department of Mechanical Engineering, University of California Riverside, Riverside, California 92521}
\author{Shane Revel} 
\affiliation{Department of Materials Science and Engineering, University of California Riverside, Riverside, California 92521}
\author{Sk Tahmid Shahriar} 
\affiliation{Department of Mechanical Engineering, University of California Riverside, Riverside, California 92521}
\author{Tamar S. Mentzel}	
\email{tamarm@ucr.edu}
\affiliation{Department of Mechanical Engineering, University of California Riverside, Riverside, California 92521}

\begin{document}
\singlespacing
\maketitle

\begin{multicols}{2}

\textbf{Colloidal nanocrystal quantum dots (QD) enable the bottom-up assembly of designer solids.\cite{ACSNano2015} Among the multitudinous applications of QD solids, there has been great success in exploiting the tunable optical properties for LED displays \cite{NaturePhotonics2018, NaturePhotonics2024, NaturePhotonics2020}, lighting, bioimaging and diagnostics.\cite{AnnualReview2013} Applications dependent on electrical properties such as solar cells \cite{NatureEnergy2024, NaturePhotonics2022, EnergyEnvMat2023}, photodetectors \cite{NaturePhotonics2024PD}, and transistors have fallen short of their full potential because of poor control over electrical properties, and some of applications with the most promise for novelty, such as a solid-state quantum simulator for quantum computation\cite{NatureNano2024, LevitovCharge1999} and spintronics, are stagnant. Lack of clarity on the charge transport mechanism has been a significant barrier to progress,\cite{NatureCommuncations2020} particularly as numerous sources of disorder are present.\cite{Romero:PRL, Talapin, Phillipe} In this work, we make advancements in a nano-patterning technique\cite{Mentzel2012} to fabricate a 70-nm wide QD solid that is also free of several sources of structural defects. Owing to the small size and structural integrity, we isolate the charge dynamics of a single conductance channel within a percolation network. We tune parameters to measure $\sim$10 channels, and with a time-resolved measurement, we find conductance noise that exceeds 100$\%$ of the average current. From observation of the long-time dynamics of the charge transport, including random telegraph noise, colored noise and attractor states, we model the transport with stochastic quasi-one-dimensional percolation paths. With this insight into the charge transport of QD solids unimpeded by structural defects, we provide a path for the rational design of a QD solid with electrical properties that reflect the underlying tunable, periodic potential.}

Colloidal semiconductor nanocrystal quantum dots (QD) form the basis for a designer QD solid owing to their tunable size, shape, composition and packing. In spite of their great progress as optical absorbers and emitters in LED displays and biological imaging and diagnostics, their performance in optoelectronic and electronic devices has suffered from poorly controlled electrical properties. Moreover, QD solids hold promise as the basis of a quantum simulator. Long-range Coulomb interactions in the QD solid are the basis for predicted unconventional charge and spin transport properties with application in quantum computation and spintronics. The primary obstacle to achieving these advancements is disorder. Among the sources of disorder are clustering;\cite{Mentzel2012} cracking; grain boundaries; variations in the tunnel barriers between nanocrystals; disorder in site energy;\cite{Mentzel2008} and unpassivated charge traps. \cite{almeida_charge_2016, vale_charge_2025} Disorder obscures not only the underlying periodic potential upon which these applications depend, but also challenges our ability to build a coherent model to describe the charge transport mechanism beyond reporting a variety of signatures of disorder. A plethora of models of disordered transport have been reported: variable range hopping, with and without a Coulomb gap;\cite{Romero:PRL, Talapin, Phillipe}  a simply activated model;\cite{Mentzel2008} space-charge-limited current and band transport.\cite{NatureNano2011, AdvElecMat2024} Theoretical models include phonon-assisted small polaron hopping;\cite{NatureCommuncations2020} mapping onto interface dynamics;\cite{PhysRevLetters1993} and mapping onto a frustrated antiferromagnetic spin model.\cite{LevitovCharge1999} Measurable signatures of disorder give information about the average electronic behavior from which it is difficult to arrive at conclusive insights into the nanoscopic charge dynamics. This is exacerbated by techniques employed to enhance the tunnel coupling, and thus enhance the conductivity, which often generate defects and increase packing disorder. Direct insight into the charge mechanism at a nanoscale is essential to improving fabrication and realizing a QD solid with a truly periodic potential.

Here, we nanopattern a QD solid comprised of \textit{n}-butylamine-capped PbS nanocrystals as the active region of a field-effect transistor that eliminates clustering, cracking and grain boundaries while still having a measurable electrical current. The elimination of clusters and cracks increases the conductivity by a factor of $\sim$190. For QD solids less than 70-nm wide, we isolate $\sim$10 conducting channels and find unusually large conductance noise, and the noise grows as the conductance increases with applied bias, gate voltage and temperature. By tuning the number of active channels, we are able to measure conductance fluctuations in a single channel and to observe trapping dynamics of a single charge. From the stochastic behavior of these quasi-one-dimensional percolation paths, we learn how the measurable electrical properties arise from the nanoscale electronic behavior. These insights provide a roadmap for achieving a strongly coupled, periodic, defect-free QD solid. Our nano-patterned QD solids provide a platform for investigating how packing and trapping affect electrical properties not just in our QD solid but to guide developments in synthesis and fabrication. 

\subsection{Conductance noise in a nano-patterned, structurally ordered quantum dot solid}
\begin{figure}[H]
\setlength{\unitlength}{1cm}
\begin{center}
\includegraphics[width=16cm, keepaspectratio=true]{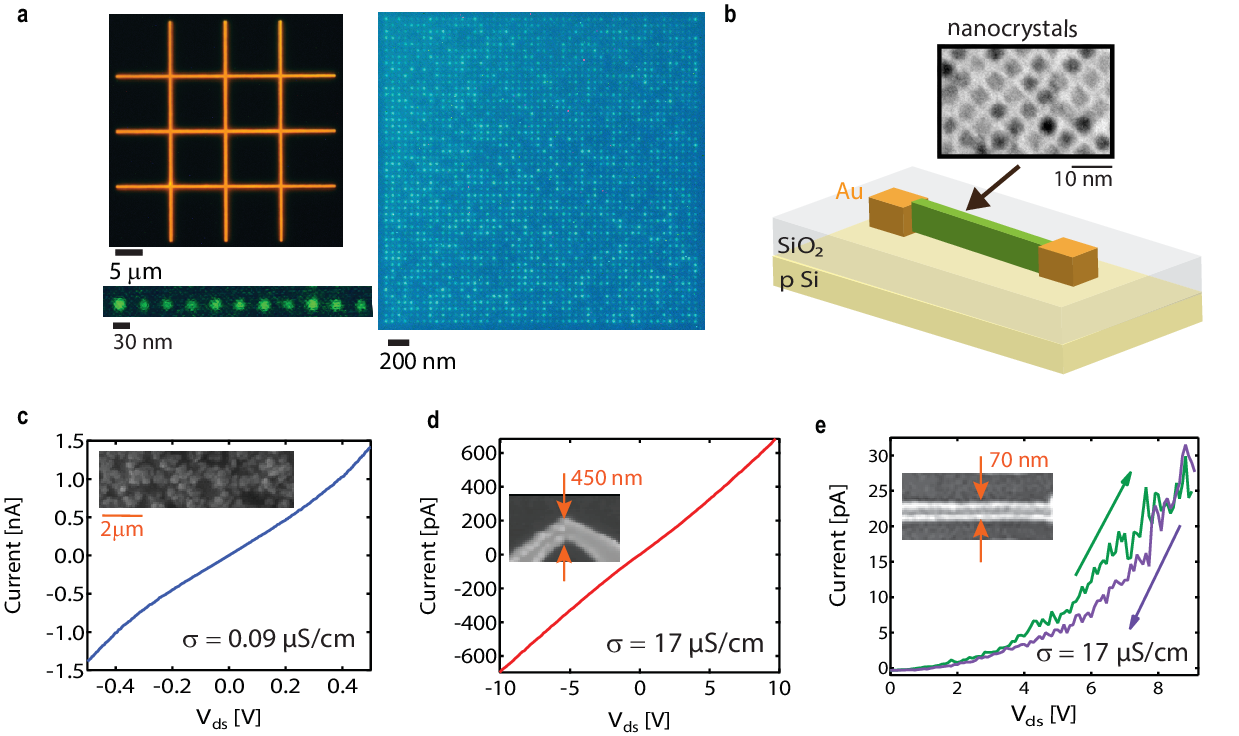}
\end{center}
\caption{(a) Films of semiconductor nanocrystals patterned with nanoscale dimensions and free of structural defects such as grain boundaries and cracks. The films are comprised of Zn$_{0.5}$Cd$_{0.5}$Se$-$Zn$_{0.5}$Cd$_{0.5}$S core$-$shell nanocrystals (cross-hatch pattern) and CdSe (dot pattern). The blue dot pattern is false color to improve visibility. (b) $\textit{n}-$butylamine-capped PbS nanocrystals form the active region of an inverted field-effect transistor. Transmission electron micrograph shows the close$-$packed array. (c)-(e) IV curves of $\textit{n}-$butylamine-capped PbS nanocrystals that are drop$-$cast and nanopatterned to 450$-$nm wide and 70$-$nm wide at its narrowest point respectively. Insets are scanning electron micrographs showing that nanopatterning results in arrays that are structurally continuous, in contrast to drop$-$cast films. With the elimination of clustering, the zero$-$bias conductance increases by a factor of $\sim$180. For sufficiently narrow channels, current noise arises. The gold electrodes on each side of the film are not shown.}
\label{fig:Size}
\end{figure}

\begin{figure}
\setlength{\unitlength}{1cm}
\begin{center}
\includegraphics[width=16cm, keepaspectratio=true]
{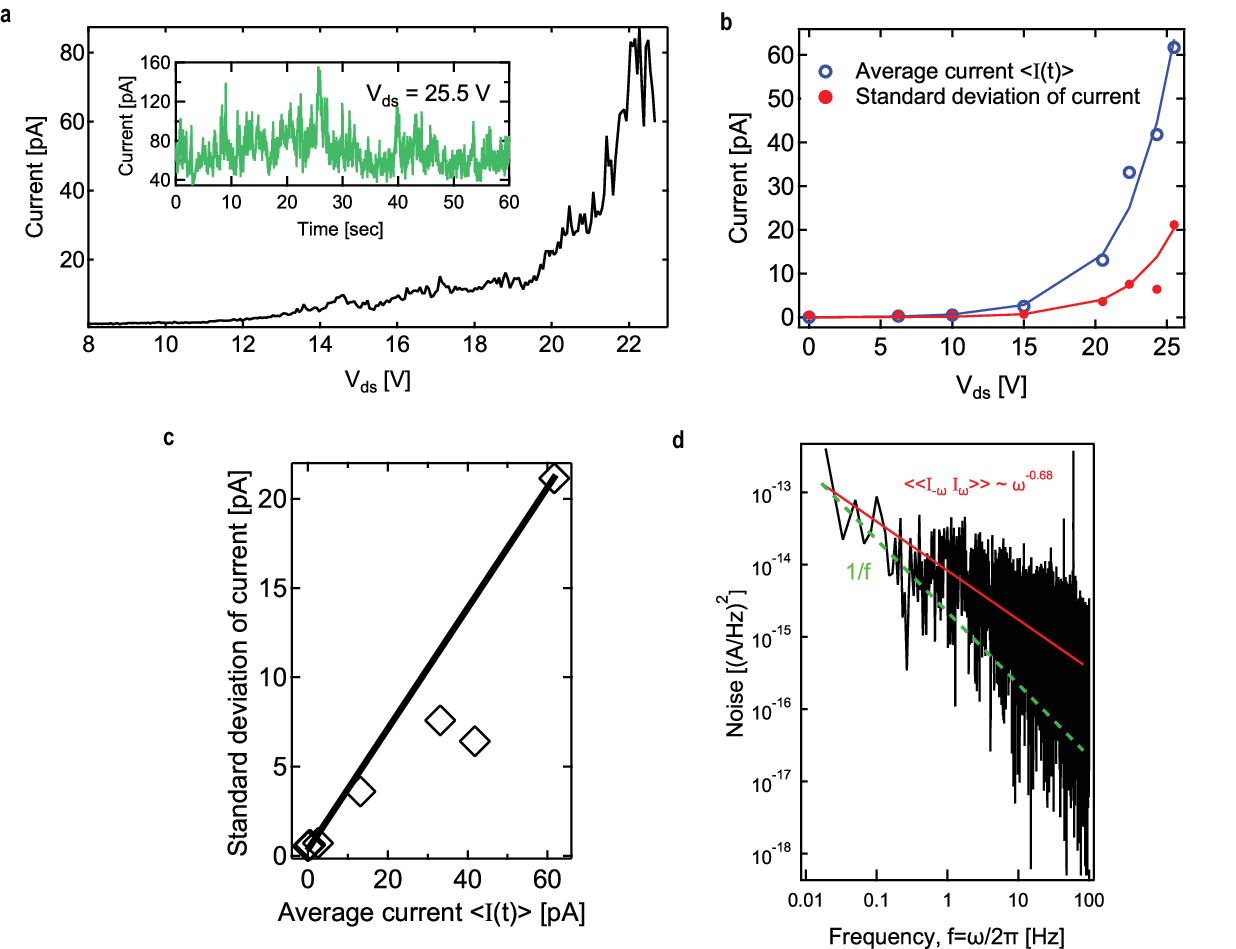}
\end{center}
\caption{(a) Current versus drain-source voltage in the nanocrystal array displayed in Fig. \fig{Size}(e).  (inset) Current versus time with a bandwidth of 0.0002 - 1 Hz at V$_{ds}$=25.5 V. (b) The average current (blue) and the standard deviation of the current (red) versus drain-source voltage.  Lines are fits to an exponential in V$_{ds}$. (c) Current noise varies proportionally to the average current, an indication that conductance fluctuations give rise to the noise. (d) Fourier transform of the noise at V$_{ds}$=13 V. The noise fits to a power law as shown in red. The green dashed line shows 1$/$f noise for contrast.}
\label{fig:Vds}
\end{figure}

We nanopattern quantum dot solids of PbS nanocrystals that are free of three sources of structural disorder commonly found in these QD solids: cracking, clustering and grain boundaries. To probe the charge dynamics in a nanocrystal quantum dot solid that is free of structural defects, we begin by synthesizing PbS  nanocrystals that are 4.5$\pm$0.2 nm in diameter by high-temperature pyrolysis of Pb and S precursors according to details provided in the Methods section.\cite{Steckel,Synthesis} We exchange the native oleic-acid ligand ($\sim$1.8-nm long) on the surface of the nanocrystal for \textit{n}-butylamine ($\sim$0.6-nm long) while the nanocrystals are still in solution. As the ligand forms an electronic tunneling barrier between nanocrystals in the solid form, the exchange for a shorter ligand increases the electrical conductivity of the solid. By performing the exchange in solution as opposed to commonly doing so once the nanocrystals are formed into a solid, we avoid the cracks generated by the latter. We drop-cast the nanocrystals from solution onto a Si/SiO$_{2}$ substrate, forming a film that is 2 $\mu$m long, 800 $\mu$m wide and approximately 100-nm thick. Clusters develop in the film as shown in a scanning electron micrograph (SEM) in Fig.\fig{Size}c, and the zero-bias conductivity is 0.09 $\mu$S$/$cm. To prevent clustering and to improve the ordering of the solid, we assemble the nanocrystals into nanoscale patterns that are smaller than the characteristic size of a cluster, $\sim$0.5-1 $\mu$m. We use a technique we previously developed entailing electron-beam lithography to make nanoscale trenches in PMMA, followed by deposition of nanocrystals and lift-off.\cite{Mentzel2012} We demonstrate nano-patterns with a variety of types of nanocrystals, as shown in Fig.\fig{Size}a. We nano-pattern an array that is 350-nm wide at its narrowest point, the region that limits the electrical conductance, and 130-nm thick.(Fig.\fig{Size}d) The SEM image shows that it is structurally continuous. We find from the IV characteristic that the zero-bias conductivity is 17 $\mu$S$/$cm, an increase by a factor of 180 by eliminating the clustering found in drop-cast films of the same nanocrystal type. Transmission Electron Micrography (Fig.\fig{Size}b) shows that the nanocrystals assemble into a rhombohedral close-packed structure. When we survey the packing over a larger scale (Supplementary Fig. 1), we find grain boundaries with a typical grain size in the range of 100-200 nm.  To further improve the ordering of the nanocrystal arrays, we eliminate grain boundaries by decreasing the width of the array to 70 nm, below the typical grain size.(Fig.\fig{Size}e) With the elimination of structural defects from cracking, clustering and grain boundaries in an array $\sim$10 nanocrystals wide and $\sim$4 nanocrystals thick, we find noise in the IV characteristic. Contrary to shot noise, the magnitude of the noise increases with increasing current.  As the drain-source voltage (V$_{ds}$) is swept from 0 to 10 V and then back down to 0 V, the noise pattern approximately reproduces itself as a function of \Vds. We set \Vds = 25.5 V and measure the current with time for one minute with a bandwidth of 0.0002 - 1 Hz.(Fig.\fig{Vds}a(inset)) We histogram the noise traces and fit the histogram to a Gaussian distribution from which we take the mean to be the average current and take the full-width at half maximum to be the noise.(Supplementary Fig. 2)  As one example, at \Vds = 25.5 V, the average current is 61.7 pA and the noise 21.1 pA, so the noise is 34$\%$ of the average current. With in this one-minute time trace, we measure a current as high 147 pA, making the fluctuations more than 100$\%$ of the average current of 61.7 pA. We repeat this for various values of \Vds\space and plot the results in Figs.\fig{Vds}b,c. We find that the standard deviation of the current varies approximately linearly with the average current. The proportionality between the current and the noise indicates that this is conductance noise, namely variations in the conductance of the channel, rather than charge fluctuations. The latter should see a reduction in the noise with increasing signal. The average current and the noise vary exponentially with \Vds.(Fig.\fig{Vds}b) As the applied voltage serves to lower the potential barrier height set up by the surface ligands, this dependence on \Vds\space is consistent with charge transport via tunneling through the barrier. We perform a time-dependent current measurement with a wider bandwidth of DC to 10 kHz at \Vds = 13 V, and take its Fourier transform.(Fig.\fig{Vds}d) The noise $<<I_{-\omega}I_{\omega}>>$ fits to a power law of $\sim\omega^{-0.68}$ where $\omega=2\pi f$. The 1$/$f dependence is shown for contrast. This power law dependence rules out the possibility that the noise in our QD solid is white noise from a Poisson process such as shot noise, or pink (1$/$f) noise.  

\subsection{Model of stochastic quasi-1D percolation paths}
To understand the origin of the conductance noise, we propose a model of N conducting channels where a stationary stochastic process gives rise to conductance fluctuations in these channels. Our nano-patterned solid comprises appproximately ten nanocrystals along the width and four nanocrystals along the height, so we expect N$\lesssim$40. Charge transport is expected to take place by majority hole carriers in lead chalcogenide nanocrystals\cite{Mentzel2008} that tunnel to neighboring nanocrystals through the potential barrier established by the ligand. While our solid is free of the aforementioned structural defects, disorder remains in the energy landscape. This remaining disorder causes variation in the hopping rates between nanocrystals as follows. (i) As the spacing between nanocrystals varies, the tunneling rates and thus the current will vary exponentially as a function of the variation in the width. (ii) Size variation in the nanocrystal, even as little as $\sim3\%$, causes a broadening by $\sim$20 meV\cite{Weidman2014} in the energy of the PbS 1S$_{h}$ states. The phonon absorption required for this inelastic tunneling will vary depending on the initial and final energy states. (iii) Charge trapping on the surface of the nanocrystal or in the ligand gives rise to fluctuations in the electrostatic environment. Owing to the long charge screening length,$\sim8$ nm\cite{Mentzel2008} nearby nanocrystals experience a random, time-varying chemical potential. This varying chemical potential will cause variations in the former two components of the inelastic tunneling rate by causing an effective change in the tunnel barrier height or a further misalignment of energy levels in neighboring nanocrystals. 
We model our QD solid as a network of resistors between localized sites, as in a percolation network. \cite{MathematicalCambridge1957} As the charge carriers in the QD solid will find the lowest resistance path, the current will be dominated by those paths that have comparably low resistance. The conductance of any single channel is set by the largest resistor, i.e. the rate-limiting resistor, in the channel.\cite{Ralls} When a trap state in one of these channels is filled, it can increase the resistance of a point in the channel by affecting the barrier height or the degree of misalignment of energy levels in neighboring nanocrystals. The channel conductance will change accordingly.  This may result in the channel turning "off," meaning the current is no longer measurable compared to other channels, or the conductance might shift to a lower but still measurable value.(Fig.\fig{Model})

When a voltage bias is applied such as in Figs.\fig{Size}e,\fig{Vds}, we see that it primarily lowers tunnel barrier heights driving the exponential increase in measurable channels.  As more channels become measurable, there is a concomitant increase in noise associated with charges being trapped and released. We can understand the reproducibility found in Fig.\fig{Size}e as follows. As the applied voltage affects the alignment of energy levels, bringing any two levels further in or further out of alignment, the values of \Vds\space for which we see a reproducibility in the noise peaks, we expect that two levels are brought into resonance or near resonance. This can be the energy state in the nanocrystal becoming resonant with a trap state so the charge is released and a channel opens, or neighboring nanocrystals' 1S$_{h}$ states are brought into resonance increasing the tunneling rate, and opening a new channel, as it no longer requires phonon absorption. 

\subsection{L\'{e}vy statistics and attractor states in time-dependent current measurements}
To further explore the noise in the conducting channels, we measure the time dependence of the current in response to a voltage step of \Vds= 40 V for 10 minutes at a temperature of 77 K.(\fig{Levy}c,d) The current can be described by a random telegraph signal (RTS). In our model, this is consistent with a single conductance channel that fluctuates between on and off states as the hole carrier is captured and released by a nearby trap state. As long as successive trapping events are uncorrelated, the process can be characterized by the off times $\tau_{off}$. We histogram the values of $\tau_{off}$ and find it fits to a power law with a long tail, $p(\tau_{off})=a/\tau_{off}^{1.1}$. (Fig.\fig{Levy}e) This form matches the probability density function for a L\'{e}vy distribution:
\begin{equation}
p(\tau_{off} >> \tau_{0})=a/\tau^{1+\mu},\hspace{0.7cm} 0\leq\mu\leq1
\label{eq:probdist}
\end{equation}
$\tau_{0}$ is a shorter time scale over which the event takes place during $\tau_{on}$, such as a unit of charge transmission. Consistent with this expression, we find $\mu$ = 0.1. When we repeat this measurement on similarly prepared devices, we find values of $\mu$ that range from 0.1 to 0.6. This long-tail distribution is characterized by a wide distribution of time scales, namely a wide distribution of times that a charge remains trapped owing to varying trap depths. \cite{PhysicalReviewB2005} This probability distribution has notably been found to describe the duration of dark states in fluorescence intermittency known as blinking.\cite{PhysRevLetters2003} In the latter, charge trapping is also thought to be the mechanism giving rise to these statistics. By increasing the measurement time, the probability of finding a charge in a long-lived trap increases, as seen in the longer values of $\tau_{off}$ at later times. 

The temperature is increased to 295 K and the voltage is stepped to \Vds = 13 V for 100 minutes.(Fig.\fig{Levy}a) We observe a slowly decaying current transient, and the noise decays proportionally with the current, consistent with the proportional relationship found previously. The rate of decay of the current fits to a power law $I(t) = I_{0} t^{-0.46}$. Based on prior theoretical work on L\'{e}vy distributions in charge transport, the average current was found to follow a power law decay of 
\begin{equation}
I(t) = I_{0} t^{1+\mu},\hspace{0.7cm} 0\leq\mu\leq1
\label{eq:currentdecay}
\end{equation}
We find $\mu$ = 0.54 in our current transient. This decay can be understood to arise from the same mechanism as the increasing value of $\tau_{off}$ in the RTS. With increasing temperature, more hole carriers are activated and higher energy phonons are available for inelastic tunneling. As such, the RTS of a fluctuating individual channel grows to a current signal comprised of several channels opening and closing with time, on the order of one to ten channels based on the dimensions of the array. The longer the measurement time, the higher the likelihood that a charge carrier is trapped in a long-lived trap state, thereby turning off a channel. As the integral of the current over time in equation \eqn{currentdecay} diverges, this is a true drain-source current rather than a displacement current.

We confirm the model by which charge is transmitted via a stochastic process given by a L\'{e}vy distrubtion by  exploring the net charge transmitted over time and the frequency spectrum. Whereas for a Poisson distribution there is a mean time $\tau$ and one expects the total transmitted charge $<Q>$ to grow linearly with time, in our L\'{e}vy distribution the mean off-time diverges so that the total transmitted charge grows sublinearly with time as $<Q>\sim t^{\mu}$ with $\mu$ = 0.72, as found in Fig.\fig{Levy}f. The distribution in equation \eqn{probdist} leads to a frequency spectrum for both the current and the noise $<I_{\omega}>\sim<<I_{-\omega}I_{\omega}>>\sim\omega^{-\mu}$ where $\mu\sim$ 0.72 for the current and 0.68 for the noise. \cite{bardou_levy_2002} (Fig.\fig{Levy}b)

Up until this point in our analysis, we have found consistency between the stochastic process observed in the charge transport behavior and that observed in nanocrystal fluorescence intermittency, both described by long-tailed L\'{e}vy statistics. A distinction is that reports of fluorescence intermittency find that the ensemble of nanocrystals are nonstationary-- aging is reported. \cite{PhysRevLetters2003} When we increase the number of channels that we measure such as in our 350-nm wide nano-patterned QD solids, we find a deterministic response to an applied voltage. The noise is eliminated, indicative of a stationary system-- no signs of aging.  When we extend our observation time of the current response to a voltage step to beyond the 10-100 minutes observed thus far, we find new dynamical properties as shown in Figure\fig{Levy}g. After the initial decay in the current, it recovers abruptly, shown as the recovery to the initial values of the current and noise, regime I. We also find a repeat in the long-tailed decay to regime II. This followed by discrete changes to regime III and a discrete recovery to regime II followed by a third cycle of a long-tailed decay. To understand these discrete changes, we have to expand our model from individual channels to percolation networks comprised of quasi-individual channels.  Owing to the long-range Coulomb interactions in the QD solid, a single trapped charge can electrostatically change the conductance of more than one channel at a time, therefore causing several channels to turn on or to turn off at once. The reproducibility of these regimes can be thought of as attractor states-- states that the system will remain in despite minor perturbations or that the system will return to when larger perturbations take place. \cite{elze_attractor_2009} Despite a nonstationary response with diverging time scales, the overall system remains stationary.

\subsection{Temperature and gate voltage tune the system properties}
We measure the temperature dependence of the conductance starting with at 4 K through 308 K.(Fig.\fig{TempGate}a) Approximately one hour elapses between measurements at each temperature. There is a general trend of thermally activated conductance and noise. Below 200 K, the measured noise is dominated by the current amplifier.  At approximately 200 K, the measured noise becomes dominated by conductance noise. We can understand the thermal activation of the conductance from thermally activating charge carriers and phonon-assistance of tunneling, as found previously.\cite{Mentzel2008} Notably, the conductance noise appears as a function of temperature and then at a sufficiently high temperature for the noise to appear, the noise can then grow as a function of applied bias. This is consistent with the required thermal energy for releasing carriers from traps. There is a monotonically increasing conductance as a function of temperature between 4 to 160 K, followed by a discrete switch to a lower conductance at 200 K. From 200 to 270 K, the conductance and the noise grows monotonically with temperature. From 270 to 308 K, we find a monotonically decreasing conductance consistent with the decay in the average current found at values close to room temperature. Overall, the noise grows with temperature. Even when stochastic trapping process takes place, causing average current to decline, we still see an increase in noise with T because more carriers are activated. At 295 K, we vary the gate voltage and confirm from the increasing current with negative values of \Vg\space that the majority carriers are holes.(Fig.\fig{TempGate}b) As we induce more hole carriers, we find a trend of increasing conductance and noise, with some discrete switches as found previously, such as between \Vg\space  of -27 and -36 V.  In the inset of Fig.\fig{TempGate}b, we show that the relative noise, namely noise divided by average current grows with increasingly negative gate voltage. This differs from the trend of L\'{e}vy statistics found earlier in which the noise grows proportionally with average current. The difference between these two cases is that earlier we observed the stochastic behavior after a step in \Vds. The voltage bias serves to increase the conductance by lowering the energy of states down field thereby decreasing the activation energy to generate charge carriers; decreasing the energy required for inelastic tunneling; and decreasing potential barriers.  In contrast, with the gate voltage, charge carriers are generated without the other changes to the tunneling rate. As such we find the noise increases faster than the average current. 

\subsection{Discussion}
With the elimination of three sources of structural defects, namely clustering, cracking and grain boundaries, we gain insight into the charge transport mechanism intrinsic to the nanocrystals. We find transport is dictated by phonon-assisted tunneling between nanocrystals that is dynamically modulated by charge trapping events. Three sources of disorder remain: disorder in site energy of the nanocrystals; variation in tunneling coupling between nanocrystals; and charge trapping. With the nano-patterned QD solid reported herein that enables direct observation of how these sources of disorder contribute to transport, it is possible to make advancements to synthesis and fabrication addressing these remaining sources of disorder and to test the improvements. Enhanced coupling between nanocrystals is expected to overcome the disorder in site energy and variations in tunnel rates\cite{abelson_high-mobility_2022, choi_bandlike_2012}, which will give rise to uniform conductance across QD solids $<$70-nm wide. Moreover, further passivation of trap states\cite{ChemistryofMaterials2019, EnergyEnvMat2023, NanoLetter2016, ACSNano2024} will resolve the stochastic conductance fluctuations. With these modifications integrated into a nano-patterned QD solid, there is a path toward realizing a designer QD solid with periodic potentials given by the packing of the nanocrystals. This would enable investigation of the predicted collective charge and spin properties as well as optoelectronic devices with optimized charge transport efficiencies.

\begin{figure}[H]
\setlength{\unitlength}{1cm}
\begin{center}
\includegraphics[width=16cm, keepaspectratio=true]
{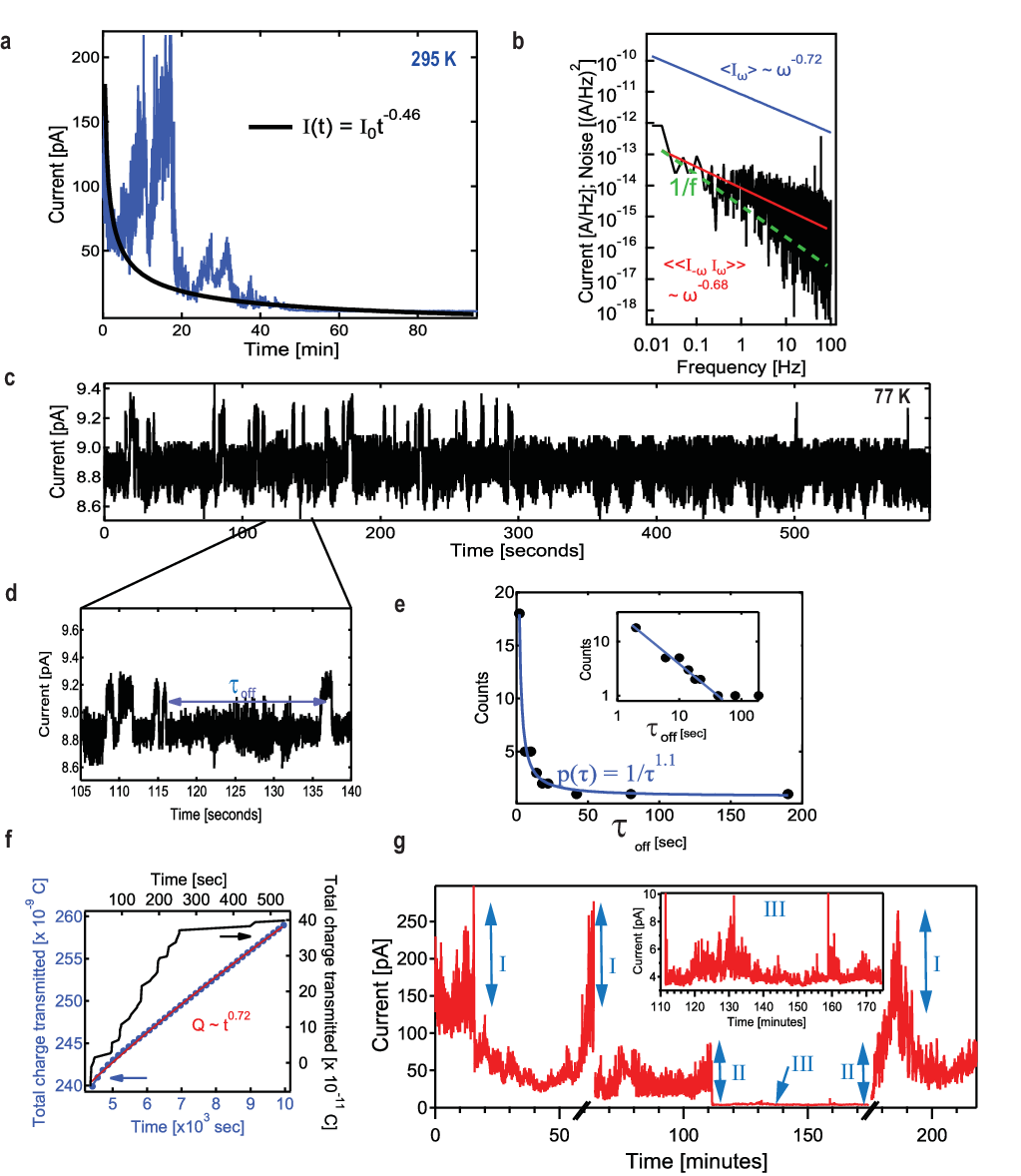}
\end{center}
\caption{(a) Current versus time at a temperature of 295 K with a voltage bias of $V_{ds}=13 V$. The black line is a fit to a power law decay of the current transient. (b) The Fourier transform of the current decay from panel a (blue line) fits to a power law with a comparable power as the Fourier transform of the current noise. (c) Current versus time at a temperature of 77 K with a voltage bias of $V_{ds}=40 V$. (d) Magnification of the signal in panel c between 105 and 140 seconds. The waiting time between events is $\tau_{off}$.  (e) Histogram of the waiting times $\tau_{off}$ from panel c fit to $p(\tau)=a/\tau^{1+\mu}$ where $\mu=0.098\pm0.12$. The inset is the histogram on a log-log plot. (f) Total charge transmitted over time, $<Q(t)>$, calculated from the current in panel a (blue circles) and in panel c (black line). Red line is a fit to $<Q(t)>\propto t^{\mu}$. (g) Current versus time at temperature of 295 K for an even longer time than in panel a. We observe three regimes with different values of the average current. The inset is a magnification of regime III.}
\label{fig:Levy}
\end{figure}

\begin{figure}[H]
\setlength{\unitlength}{1cm}
\begin{center}
\includegraphics[width=16cm, keepaspectratio=true]
{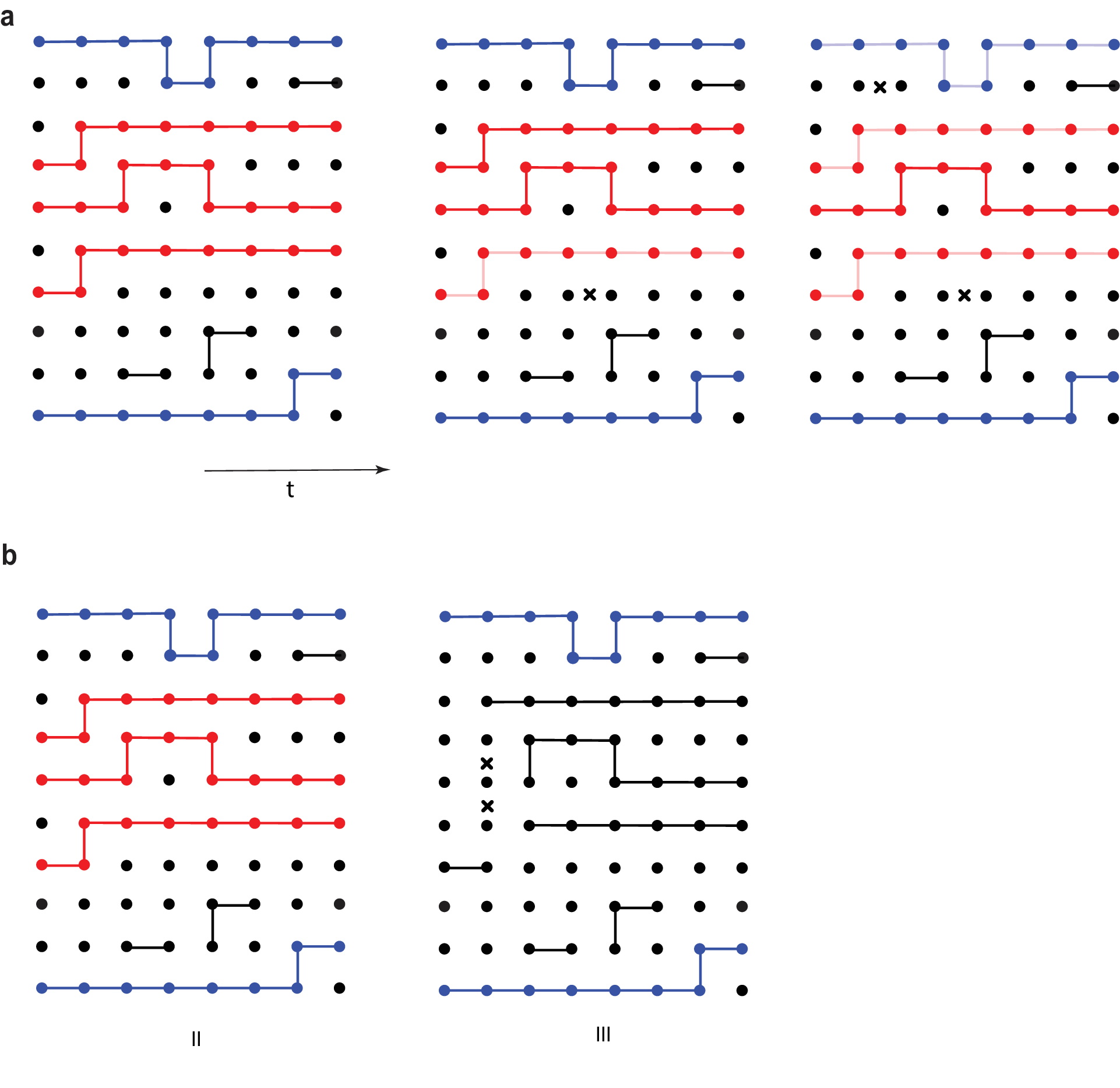}
\end{center}
\caption{(a) Percolation paths representing the mechanism of the current transients. The red paths represent the conduction channels that dominate the current, and the blue paths represent connected paths with slower rates of charge transmission. As traps fill over time, represented by x, the long screening length cause reductions in the tunnel rates of nearby channels, shown by the diminished color. The rate of the dominant red paths gradually diminishes. (b) Percolation paths representing the discrete shift to an attractor state of lower conductance. Filling of traps sufficiently nearby to several dominant red channels diminish the rate below the resolution limit. The blue paths become dominant.}
\label{fig:Model}
\end{figure}

\begin{figure}
\setlength{\unitlength}{1cm}
\begin{center}
\includegraphics[width=16cm, keepaspectratio=true]
{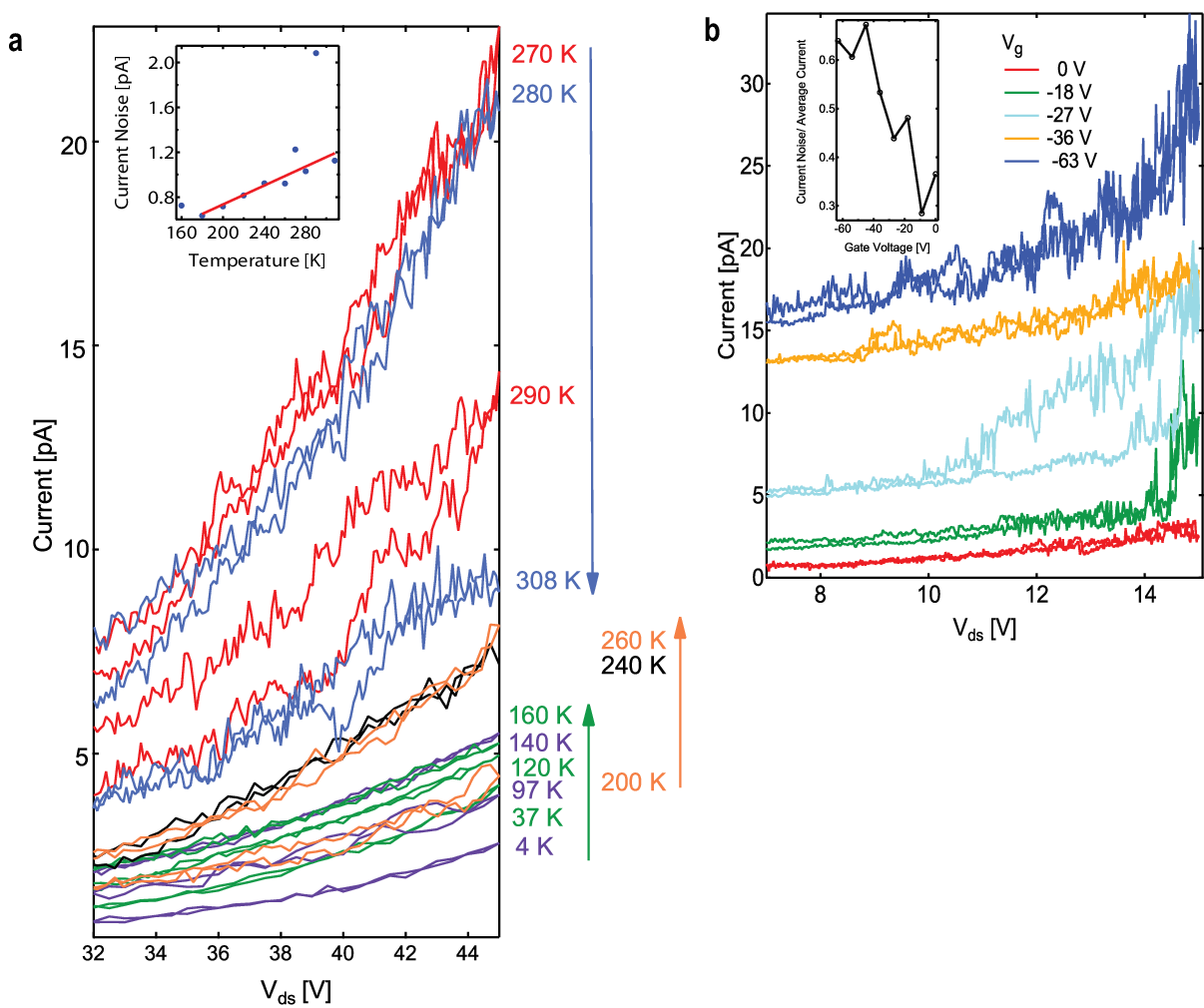}
\end{center}
\caption{(a) Current versus source-drain voltage as a function of temperature. The inset shows the current noise versus temperature. The noise calculated from time traces of the current at V$_{ds}$ = 44 V for ten minutes with a bandwidth of 0.002 - 100 Hz. (b) Current versus source-drain voltage as a function of gate voltage. The inset shows that the relative noise as a function of negative gate voltage.}
\label{fig:TempGate}
\end{figure}

The ability to measure charge transport dynamics, particularly the charge fluctuations in time, in an individual conducting channel gives rise to the question at the core of statistical physics, namely what is the relation between the stochastic behavior of an individual channel and the an ensemble. The latter is important to understand as it reflects what we find in a larger scale QD solid. This question is particularly salient as systems governed by a L\'{e}vy distribution have been reported to be nonergodic, that is the long time average of a property is not equivalent to the ensemble average of that property. In single nanocrystals, fluorescence intermittency, also known as blinking is observed.\cite{PhysRevLetters2003, NatureMaterials2013, ChemPhys2019} Long time averages of the fluorescence signal vary significantly from nanocrystal to nanocrystal indicative of nonergodicity. The fluorescence signal is described by a L\'{e}vy distribution owing to the wide distribution of trapping lifetimes. The nonstationarity of the fluorescence is consistent with a diverging mean for a Levy distribution. There are both similarities and important differences to the Levy statistics we find herein. In our long-time transport measurements we find charge dynamics represented by a L\'{e}vy distribution in intermediate times. However for the full duration of the measurement, the system resets and follows deterministic behavior over a long enough time.(Fig.\fig{Levy}g) This affirms the ergodic hypothesis and is indicative of a stationary system. In the interest of building up a QD solid from constituent nanocrystals, it is important to differentiate how trapping presents in individual nanocrystals versus in individual conductance channels with in the matrix of the QD solid.

There are other examples of charge transport that are dominated by trapping dynamics as we find here. In disordered systems like amorphous semiconductors, dispersive transport has been reported describing how charges hop between localized states.\cite{PhysRevB1975, PhysRevLetters1981, SolidStateComm1981} Levy distributions have been found to describe the time dynamics. It is noteworthy that our QD solid differs from dispersive transport as in the latter the current grows with time as traps are filled. In contrast, we find a decaying current as traps are filled, followed by a recovery as charges are realeased.(Fig.\fig{Levy}f) In all-inorganic QD solids,\cite{NatureNano2011, AdvElecMat2024} trapping is also reported to dictate the transport mechanism. At low temperatures, disordered localized transport is reported owing to the presence of trap states. At high temperatures, the mobility is found to decrease with increasing temperature, which is understood as a signature of band-like transport, owing to charges all being released from traps. This contrasts with our system where trapping dynamics are evident at temperatures as high as 308 K. Questions remain how to interpret mobility variations with temperatures. Measuring mobility relies on reliable control of charge density, which can be affected by traps. Our nano-patterned QD solid provides a platform for a direct measure of single charge trapping dynamics, which would resolve this question.  Finally, as our nano-patterned QD solid serves as a tunable manifestation of Levy statistics, it may provide a testbed for exploring Levy statistics in other fields including chaotic systems, glassy dynamics and rare financial events.\cite{gomes_random_2016, boulle_statistical_2016, applebaum_levy_2004, ni_threshold_2023}

\end{multicols}


\newpage

\newpage
\bibliography{Bib_Noise_1}



\end{document}